 \documentclass[12,preprint]{aastex}

\newcommand\etal{{\rm et~al\/}.}

\def\alwaysmath#1{\ifmmode {#1}  
                  \else {$#1\mkern-5mu$} \fi} 
\newcommand\km{\alwaysmath{\,{\rm  km}}} 
 
\newcommand\cw{\alwaysmath{\,{\rm  cm}^{-1}}} 
 
\newcommand\kms{\alwaysmath{\,\km\,{\rm s}^{-1}}} 
 
\newcommand\angstrom{\alwaysmath{\,{\rm \AA}}} 
\newcommand\angs{\angstrom}

\newcommand\gyr{\alwaysmath{\,{\rm  Gyr}}}

\newcommand\kel{\alwaysmath{\,{\rm  K}}}

\newcommand\astron{Astronomy}
\newcommand\dept{Dept.~}

\newcommand\das{\dept of \astron}

\newcommand\uof{University of~}
\newcommand\aadv{Astrophysical Advances}

\newcommand\uc{\uof California}

\newcommand\uva{\uof Virginia}

\newcommand\gsfc{Goddard Space Flight Center}

\newcommand\lick{UCO/Lick Observatory}

\newcommand\nasa{National Aeronautics and Space Administration (NASA)}

\newcommand\stsci{Space Telescope Science Institute}
\newcommand\ackhst{Based on observations obtained with the Hubble Space 
Telescope of \stsci, under contract with the \nasa}

\newcommand\acklick{Based on observations obtained with the Shane Telescope at 
Mt.\ Hamilton, \lick}

\newcommand\acksim{This work has made use of the SIMBAD database, operated at 
CDS, Strasbourg, France}

\newcommand\bd{
  Emergent-IT Inc.\ and \gsfc, Greenbelt, MD 20771; Ben.Dorman@gsfc.nasa.gov}
\newcommand\rtr{
   \das, \uva, P.O. Box 3818, Charlottesville, VA  22903-0818; 
rtr@veris.astro.virginia.edu}
\newcommand\rcp{
   \lick, \das, \uc, Santa Cruz, CA 95064, and 
\aadv, Palo Alto, CA  94301; peterson@ucolick.org}
\newcommand\gapprox{\ga} 
\newcommand\lapprox{\la} 

\newcommand\ilc{i}
\newcommand\iilc{ii}
\newcommand\msun{\alwaysmath{\,M_\odot}}

\newcommand\logr{\alwaysmath{ \log\,R }} 
\newcommand\logg{\alwaysmath{ \log\,g }}

\newcommand\teff{\alwaysmath{T_{{\rm eff}}}}

\newcommand\mgb{\alwaysmath{{\rm Mg} b}}

\newcommand\feh{{\rm [Fe/H]}}
\newcommand\life{\alwaysmath{\rm [Li/Fe]}}
\newcommand\alfe{\alwaysmath{\rm [\alpha/Fe]}}
\newcommand\ofe{\alwaysmath{\rm [O/Fe]}}
\newcommand\mgfe{\alwaysmath{\rm [Mg/Fe]}}

\def\2h#1{\alwaysmath{{\rm [#1/H]}}}
\def\2fe#1{\alwaysmath{{\rm [#1/Fe]}}}

\newcommand\oi{{\rm O\,{\textsc \ilc}}} 
\newcommand\mgi{{\rm Mg\,{\textsc \ilc}}} 
\newcommand\mgj{{\rm Mg\,{\textsc \iilc}}} 
\newcommand\sii{{\rm Si\,{\textsc \ilc}}} 
\newcommand\sij{{\rm Si\,{\textsc \iilc}}}

\newcommand\tii{{\rm Ti\,{\textsc \ilc}}}
\newcommand\tij{{\rm Ti\,{\textsc \iilc}}}

\newcommand\fei{{\rm Fe\,{\textsc \ilc}}}
\newcommand\fej{{\rm Fe\,{\textsc \iilc}}}

\newcommand\bmv{\alwaysmath{B-V}} 

\newcommand\umb{\alwaysmath{U-B}} 
 
\newcommand\bmy{\alwaysmath{b-y}}

\newcommand\Halpha{\alwaysmath{{\rm H}\alpha}}

\newcommand\vt{\alwaysmath{\chi_{\rm t}}} 
\newcommand\vsini{\alwaysmath{v \sin \ilc}}

\newcommand\hd{\alwaysmath{\rm HD\,}}
\newcommand\acena{\alwaysmath{\alpha\,{\rm Cen\,A}}}
\slugcomment{To appear in {\it The Astrophysical Journal, September 20, 2001}}

\shorttitle{UV Stellar Temperatures}
\shortauthors{Peterson \etal}

\begin{document}

\title{Modeling Mid-Ultraviolet Spectra. I. Temperatures of Metal-Poor 
Stars\altaffilmark{1,2}}

\author{
Ruth C. Peterson\altaffilmark{3}, 
Ben Dorman\altaffilmark{4}, and
Robert T. Rood\altaffilmark{5} 
}

\altaffiltext{1}{\ackhst.}

\altaffiltext{2}{\acklick.}

\altaffiltext{3}{\rcp.}

\altaffiltext{4}{\bd.}

\altaffiltext{5}{\rtr.}

\begin{abstract}
Determining the properties of remote globular clusters and elliptical galaxies 
using evolutionary population synthesis
requires a library of reliable model stellar fluxes. 
Empirical libraries are limited to spectra of stars in the solar neighborhood, 
with nearly solar abundances and abundance ratios. We report here a first step 
towards providing a flux library that includes nonsolar abundances, based on 
calculations from first principles that are calibrated empirically. Because the 
mid-ultraviolet spectrum of an old stellar system is dominated by the 
contribution from its main-sequence turnoff stars, we have started by modeling 
these. We have calculated mid-ultraviolet spectra for the Sun and nine nearby, 
near-main-sequence stars spanning metallicities from less than 1/100 solar to 
greater than solar, encompassing a range of light-element abundance 
enhancements.

We first determined temperatures of eight of the stars by analyzing optical 
echelle spectra together with the mid-ultraviolet. Both could be matched at the 
same time
only when models with no convective overshoot were adopted, 
and only when an approximate chromosphere was incorporated near the surface of 
relatively metal-rich models. Extensive modifications to mid-UV line parameters 
were also required, notably 
the manual assignment of approximate identifications for mid-UV lines missing 
from laboratory linelists. 
Without recourse to additional missing opacity, these measures suffice to 
reproduce in detail almost the entire mid-UV spectrum of solar-temperature stars 
up to one-tenth solar metallicity, and the region from 2900\angs\ to 3100\angs\ 
throughout the entire metallicity range. 
Ramifications for abundance determinations in individual metal-poor stars and 
for age-metallicity 
determinations of old stellar systems are briefly discussed, with emphasis on 
the predictive power 
of the calculations.

\end{abstract}


~~
~~
~~
~~
~~
\pagebreak

\section{Introduction}

To characterize the metallicity and age of an old stellar system, iron abundance 
\feh\ 
(the logarithm of the iron-to-hydrogen abundance ratio with respect to its solar 
value) 
and the stellar effective temperature \teff\ must be determined together for its 
main-sequence turnoff (MSTO) stars. These are  dwarfs and subgiants of 
types F and early G with $5750 \leq \teff \leq 7000\kel$. Their spectra are 
dominated by \fei\ lines,
whose strength increases as \teff\ declines. For both metallicity and age, then, 
a reliable derivation of \teff\ is necessary.

This can be problematical even for an individual star.
The photometric colors often used to derive \teff\ are sensitive to reddening by 
the interstellar medium and to the modeling of convection (Castelli, Gratton, \& 
Kurucz 1997), 
as is the more recently developed infrared flux method \citep{alo96}. 
Methods using Balmer-line profiles are reddening-independent but still very 
model-dependent \citep{mag84,fuh94}. Methods relying on the excitation 
equilibrium of \fei\ are 
reddening-independent and model-independent if lines are weak, but are 
susceptible to systematic errors in 
laboratory measurements of gf-values \citep{bla82} and to possible departures 
from local 
thermodynamic equilibrium \citep[LTE;][]{the99}.  

Currently, \teff\ values derived for individual MSTO stars show uncomfortably 
large spreads: values of 
\citet{car94} and \citet{ful00} average 100\kel\ -- 200\kel\ lower than those 
found by \citet{kin93} 
and \citet{gra96}, with those of \citet{alo96} and \citet{bla98} in between. Not 
only is overall metallicity affected, but also the ratios of the abundance with 
respect to iron of certain elements. Among these are the \life\ and \ofe\ ratios 
so critical to uncovering the nucleosynthesis history of the early universe 
\citep{spi96} and deducing the age of globular clusters from the MSTO luminosity 
\citep{vdb00}. When \teff\ alone is raised by 100\kel, modeled values of \life\ 
change by +0.09\,dex \citep{spi96}; \ofe\ changes by $-0.09$\,dex, +0.03\,dex, 
or +0.16\,dex depending on whether the high-excitation \oi\ 7775\angs\ triplet, 
the [\oi] 6300\angs\ ground-state line, or OH molecular lines near 3100\angs\ 
are used \citep{kra00}. Not surprisingly, the oxygen abundance in extremely 
metal-poor stars remains uncertain, as summarized during a workshop devoted to 
the problem \citep{kra00,bal00,pet00b,lam00}. \citet{kin93} and \citet{kra00}, 
among others, have suggested that a higher \teff\ scale for metal-poor turnoff 
stars might eliminate the systematic differences in \ofe\ seen from its 
different diagnostics.

For all these reasons, it seems important to reduce uncertainties in the 
determination of \teff\ in MSTO stars. Fitting the slope of their UV flux 
distribution might help, since the fluxes of F--G stars peak in the optical so 
that UV fluxes vary rapidly with \teff . Although fitting a UV slope is also 
reddening-dependent, reddening should be very low for the brightest metal-poor 
MSTO stars, which are typically within 50\,pc \citep[e.g.,][]{rei98}. 

Reliable calculations of MSTO mid-UV spectra would be especially valuable in 
constraining the age and metallicity of a spatially unresolved old stellar 
system such as an elliptical galaxy or extragalactic globular cluster. For the 
mid-UV flux of an old stellar population is dominated by turnoff stars 
\citep{dor93}, but the light redward of 4500\angs\ is dominated by cool giants 
in a 17\gyr\ population of solar metallicity \citep{wor94}. To date, however, 
mid-UV studies of galaxies and globulars \citep[e.g.,][]{spi97,pon98} have 
largely been empirical, comparing extragalactic spectra against observed spectra 
of individual F and G 
stars \citep{fan90}. Thus the metallicity of the extragalactic systems has 
generally been assumed to be solar, although age is very sensitive to this 
assumption \citep{hea98}. Furthermore, nonsolar abundance ratios have not been 
included, despite the fact that abundance enhancements of light elements such as 
magnesium are generally seen in metal-rich extragalactic systems 
\citep{pet76,oco76, wor92,hen99}.

Modeling the mid-UV region is difficult, 
though, because of the dramatic increase in line absorption at short 
wavelengths. Line blending in the 
solar spectrum suppresses the optical continuum blueward of $\sim$ 4500\angs\ 
\citep{kur84}, and 
becomes even more severe in the near- and mid-UV. Especially in the UV, many 
lines are ``missing'' -- although they appear as absorption features in spectra 
of solar-type stars, they lack laboratory identification, and are thus not 
matched by spectral synthesis calculations \citep{kur81}. As a result, the 
normalization of the solar UV continuum is controversial 
\citep{bal98,boe99,isr01}. 

With few exceptions \citep[e.g.,][]{hea98}, calculations of the mid-UV region 
are based on Kurucz model flux distributions. \citet{all00} compared these to 
International Ultraviolet Explorer (IUE) 
mid-UV spectra of nearby metal-poor halo stars, illustrating that \teff\ is 
generally recovered when 
parallaxes are known and reddening negligible. \citet{lot00} have used Kurucz 
fluxes as generated by \citet{lej97} to examine the effect of abundance on the 
mid-UV light of old stellar populations, but found that the UV fluxes of cool 
stars were poorly reproduced. 

This is not surprising given the known limitations of the opacity distribution 
functions on which the Kurucz flux distributions are based. They statistically 
treat lines in a list that includes ``predicted'' lines whose wavelengths are 
uncertain by $\sim$10\angs\ (because one or both energy levels have not been 
measured in the laboratory), limiting the resolution of the flux distributions 
to 10\angs. 
Moreover, they assume a solar iron abundance 
$\sim$0.15 dex higher than the value currently accepted \citep{bie91,hol95}, and 
a solar ratio of light-element abundances \alfe\ = 0. In contrast, metal-poor 
stars of the Galactic halo 
typically show \alfe\ = 0.3 \citep{whe89}, a factor-of-two enhancement in Mg, 
Si, Ca, and Ti. This affects the fit in the mid-UV, since it includes the very 
strong resonance doublet of \mgj\ at 2800\angs\ 
and a bound-free opacity edge of \mgi\ at 2512\,\AA, plus many weaker Mg, Si, 
and Ti lines. 

\section{Overview of This Work}

In this work we calculate the mid-UV spectrum line-by-line at high resolution, 
independent of the Kurucz flux distributions. Our line list does not include 
predicted lines, but rather is based on an updated laboratory line list 
augmented by manual addition of missing lines where necessary. To ensure the 
reliability of these calculations, we match as best we can the individual line 
blends observed in the mid-UV echelle spectra of real near-turnoff stars, 
spanning a wide metallicity range and including both solar and enhanced 
light-element ratios. Our purpose is to reproduce mid-UV spectra of turnoff stars 
accurately enough to determine \teff\ and to predict the composite mid-UV 
spectra of old stellar systems over their entire range of overall metallicity 
and light-element abundance ratio. 

In this first step, we describe this calibration procedure. Because the 
calibrations depend significantly on the stellar temperatures adopted, we have 
carried out new \teff\ determinations in seven metal-poor turnoff stars and one 
mildly metal-poor hotter star. They are based on simultaneously matching both 
mid-UV spectra (2280 -- 3120\angs) and the \Halpha\ profile and the strengths of 
high-excitation atomic lines in optical spectra. We thus do not rely on any 
photometry or parallax information in the \teff\ determination. Iron abundances 
and \alfe\ ratios were adopted as derived from both weak and strong lines in the 
optical spectra. To assist in filling in the mid-UV line list, we have included 
mid-UV analyses of two stars with mid-UV spectra of very high resolution, the 
Sun and \hd 128620 (= \acena), a well-studied nearby southern star with a 
greater-than-solar metallicity \citep{neu97}, for which \teff\ was taken from 
the literature. 

In fitting each mid-UV spectrum, we adopted a single continuum normalization 
constant,
the scale factor by which the observed mid-UV spectral flux was multiplied to 
match the flux predicted by theoretical calculation. 
This constant proved to agree to an average of 6\% $\pm$ 5\% with that expected 
from the ratio of stellar distance (found from the parallax) to stellar radius 
(found from the model \logg\ assuming a reasonable stellar mass). Model \umb, 
\bmv, and \bmy\  colors all agreed extremely well with those observed, 
confirming \teff\ errors of $\lapprox$50\kel.

Our mid-UV and optical echelle spectra are described in \S 3. As discussed in \S 
4, \teff\ was initially established from the optical spectra, and compared to 
the mid-UV fit. For the most metal-poor stars, this fit did not require the 
addition of missing lines, but did require the use of Kurucz models in which 
convective overshoot was turned off, in contrast to the Kurucz models in general 
use. To reproduce mid-UV line strengths and to match the pseudocontinuum of the 
mid-UV spectra of more metal-rich stars, mid-UV atomic line parameters were 
changed and missing lines were added. Transitions especially sensitive to the 
surface temperature were still poorly reproduced; this mismatch was reduced by 
adopting a mild enhancement of temperatures in shallow photospheric layers of 
the models of one-tenth solar metallicity and higher. 

The degree to which the resulting calculations match both mid-UV and optical 
spectra of turnoff stars is shown and discussed in \S 5. The match is very good 
for 2900 -- 3120\angs\ at all metallicities, and generally good throughout the 
mid-UV for metallicities at and below one-third solar. It could be improved at 
solar metallicities by obtaining additional high-resolution mid-UV spectra of a 
few well-chosen standards. 

Our \teff\ results are presented  in \S 6, where comparisons are made of model 
fluxes and colors with observations. We conclude in \S 7 with a summary and a 
discussion of the prospects for constraining age and metallicity of quiescent 
stellar systems by comparing such calculations to spectra of their integrated 
light. 

\section{Observations} 

We begin by listing the properties of the stars considered in Table 1. The first 
five columns following the HD number give the observed $V$ magnitude, colors, 
and the 
parallax from SIMBAD. The next five columns, discussed in \S 6, show comparisons 
of observed properties of each star with those calculated from the models. The 
parameters of the model used to calculate both mid-UV and optical spectra of 
each star are given in the next four columns. In all cases but \hd 128620 = 
\acena\ and the Sun, all the model parameters were redetermined here as 
described below. The final columns show \teff\ values determined previously in 
several works.

Of the eight stars whose temperatures have been redetermined here from optical 
and UV spectra, seven are metal-poor main-sequence turnoff (MSTO) stars. The 
eighth, \hd 128167, is a mildly-metal poor star lying above the halo 
main-sequence turnoff, either younger or a blue straggler. It was included to give 
temperature leverage to the mid-UV comparisons, as discussed below. Likewise the 
Sun and \acena\ were included to represent the high-metallicity end. The solar 
parameters were considered so well determined that no re-evaluation was 
necessary. The star \acena\ was not reanalyzed because no optical spectrum was 
available: it is inaccessible from the northern hemisphere. Its \teff\ and 
abundance have been determined to moderate accuracy from extensive abundance 
analyses, and its gravity as \logg\ = 4.33 $\pm$ 0.02 as found from its visual 
and radial-velocity motion about its companion \citep{chm92,neu97,pou99}. The 
iron abundance listed in Table 1 was adopted to best match its mid-UV spectrum. 

Optical spectra were obtained for all stars in Table 1 but the Sun and \acena. 
The Lick 
Observatory Shane 3m echelle spectrograph \citep{vog87} was used with a TEK 
$2048\times2048$ CCD, at a FWHM resolution of 38,000. The spectra of \hd 84937, 
\hd 94028, and \hd 106516 were kindly provided by \citet{ful00}. Exposure times 
on the 3m were typically 10min for all stars except \hd 128167 (40\,s). Complete 
spectral coverage over 3900\angs\ -- 7800\angs\ and S/N $>$ 100/pixel were 
obtained in all cases but one, although fringing compromises the accuracy of the 
data at the long-wavelength end of the range. For the one exception, \hd 114762, 
the blue spectrum was obtained in two one-hour exposures 
by Tony Misch at Lick with the coude auxiliary telescope, using the same echelle 
spectrograph configuration. For the Sun, the flux spectrum of \citet{kur84} was 
downloaded from the Kurucz web site at http://cfaku5.harvard.edu; the 
full-resolution version was matched at 150,000 resolution.

The mid-UV spectral observations of the Sun modeled here are the rocket spectra 
of \citet{kur81}, 
recorded on film. Both passes of the center of the solar disk were used. 
Gaussian broadenings of 1.5\kms\ for macroturbulence and 80,000 FWHM for 
other broadening sources provided a good match. A constant normalization scale 
factor of 1.10 was 
adopted, in keeping with the stated 10\% uncertainty of its flux calibration.

The stellar UV spectra were obtained with the Space Telescope Imaging 
Spectrograph (STIS) of Hubble Space Telescope (HST) from mid-1998 to mid-1999. 
Salient features are noted here; full details may be found from the HST archive 
at http://archive.stsci.edu. 

To obtain the mid-UV spectra of most stars, the near-UV MAMA detector was used 
with the STIS E230M grating over its default range of 
2280\angs\ -- 3120\angs\ at a FWHM resolution $\sim$25,000. \hd 106516 (= 
HR\,4657) was observed by \citet{hea98} in program \#7433 with the 
$0.2\times0.06$ aperture. Mid-UV spectra for \hd 19445, \hd 84937, \hd 94028, 
\hd 114762, \hd 184499, and \hd 201891 were acquired in snapshot program \#7402 
by \citet{pet00}, with two exposures obtained for \hd 19445, \hd 84937, and \hd 
114762. The $0.2\times0.2$ aperture was used and exposure times were 6 -- 
10\,min. Consequently, S/N is lower, but flux levels are less dependent on 
pointing. Fluxes good to 1\% for these six stars are indicated by previous 
observations of \mgj\ line profiles in each star, made using the HST Goddard 
High-Resolution Spectrograph (GHRS) and the $2\times2$ aperture in program 
\#5869 \citep{pet97}.

Mid-UV spectra of the stars \hd 128167 (=HR\,5447), from \#7433, and \hd 128620 
(= HR\,5459 =\acena), from \#7263 \citep{lin00}, were recorded in multiple 
exposures with the E230H grating. Spectral coverage is 2380\angs\ -- 3160\angs\ 
for \hd 128167 and $<$2280\angs\ -- 3160\angs\ for \hd 128620. 
A FWHM resolution of $\sim$60,000 was achieved with the $0.2 \times 0.06$ 
aperture for \hd 128167, and 
the $0.2 \times 0.05$ND aperture for \hd 128620. 
For neither star do fluxes always reproduce exactly in orders obtained in more 
than one exposure; 
differences as large as 10\% are seen. Similar uncertainties are anticipated for 
\hd 106516, where the $0.2 \times 0.06$ aperture was also used.

Rotation is detected in at least two stars and possibly in two more, from the 
extra broadening required to match line profiles. For \hd 128167, \hd 106516, 
\hd 84937, and \hd 114762, line breadths were matched in the spectral synthesis 
by adopting rotational velocities 
\vsini\ = 9, 7, 5, and 4\kms\ respectively. Thus the resolution of the mid-UV 
spectrum of 
\hd 128167 is effectively degraded to that of the stars observed at lower 
resolution, affecting the discernment of mid-UV lines. This star nonetheless has 
one of the sharpest-lined spectra among those obtained in program \#7433.

\section{Spectrum Analysis}

Spectrum analysis was based on visually matching each observed spectrum
to {\it ab initio} radiative-transfer calculations assuming static models in 
LTE. 
The program SYNTHE 
\citep{kur81}, the \citet{kur95} and \citet{cas97} grids of ATLAS9 models, and 
the Kurucz line lists 
for atomic species and molecular hydrides were downloaded from the Kurucz web 
site, where further details on each of these can be found. As described below, 
all models, including that of the Sun, were ultimately drawn from the 
\citet{cas97} grid alone. The more metal-rich ones were modified by introducing 
enhanced temperatures in the shallow layers.

The SYNTHE program accepts as input both line lists and a model atmosphere with 
temperature and other quantities tabulated for each depth, and calculates an 
array of spectra at individual emergent angles. These may then be used 
individually (as for spectra of the center of the solar disk) or convolved with 
a zero or positive rotational velocity to generate an emergent flux spectrum. In 
either case, the resulting spectrum can be broadened by Gaussians (or other 
profile shapes) to incorporate macroturbulence and instrumental broadening.

SYNTHE was modified to
run under UNIX on a Sun Ultra-30 at the University of Virginia. Mid-UV spectra 
from 2280\angs\ to 
3160\angs\ were calculated at a resolution of 330,000 (500,000 for the Sun and 
\hd 128620), and 
optical spectra at a resolution of 500,000. A Gaussian macroturbulent broadening 
of 1.5\kms\ was 
assumed for all stars as well as for the Sun. Spectra for some stars were 
rotationally broadened as listed in the last paragraph of \S 3, as was the solar 
flux spectrum, for which a mean rotational velocity of 2\kms\ was adopted. The 
spectra were then broadened by Gaussians corresponding to the values of 
instrumental resolution listed individually in \S 3.

The line lists and the source code of the opacity routines available at the 
Kurucz web site give specifics on the wide variety of opacity sources treated by 
SYNTHE. For example, the significant absorption by the 
bound-free absorption edge of \mgi\ near 2512\angs\ is treated both as line 
opacity, with the 
line list including the bound-bound transitions to $N \sim 60$, and as 
continuous opacity blueward of the bound-free edge at $\nu_o$ = 39759.8 cm$^{-2}$,
by the subroutine MG1OP. Near $\nu_o$, this routine adopts a cross section 
of 40 Mbarn with a $(\nu/\nu_o)^{-14}$ frequency dependence, based on the 
laboratory measurement of 46 $\pm$ 12 Mbarn by \citet{lom81} and the convergence 
of high-level lines. Where this formalism falls below the theoretical cross 
section, the program adopts the latter, with a cross section of 20 Mbarn and a 
$(\nu/\nu_o)^{-2.7}$ frequency dependence \citep[e.g.,][]{but93}.

For calculations redward of 5000\angs, the line list of \citet{pet93} based on 
the solar spectrum was 
adopted, with modifications due to the use here of a flux, rather than central 
intensity, spectrum for the Sun and to the incorporation of an approximate 
chromosphere in the solar model. 
This led to the deduction of larger gf-values for the strongest lines except 
where laboratory gf-values were adopted, as for low-excitation \fei\ lines with 
furnace measurements \citep{bla82}.  

Blueward of 5000\angs, the Kurucz 25 May 1998 atomic line lists 
{\it gf}\,0300.100, {\it gf}\,0400.100, and {\it gf}\,00500.100 were used. 
These differ significantly from those used in the Kurucz flux distribution 
calculations in including only atomic lines identified in the laboratory (with 
laboratory gf-values where available) -- like those of \citet{kurbel95} but with 
the addition of many lines of \fei, \fej, and other species whose upper energy 
levels were newly determined in near-UV laboratory spectra 
\citep[e.g.,][]{nav94}. Because the strengths or profiles of many lines were 
mismatched when using line parameters directly from this list, their gf-values 
and damping constants were changed where necessary, as described shortly. Since 
many lines appear in the STIS mid-UV spectra that were not modeled by lines in 
the list, such missing lines were added blueward of 3120\angs. Redward of this, 
no missing lines were added, but gf-values were changed to match stronger lines 
in the Sun in the regions depicted in Fig.\ 4 below.

The \teff, \logg, \feh, \vt, and \alfe\ values for each reanalyzed star were 
initially established from \Halpha\ and weak  lines of neutral species in the 
optical spectra, reassessed from optical lines of once-ionized species, then 
confirmed or revised from the UV. For the two most metal-poor stars, mid-UV 
spectra provided an immediate check on \teff. For the others, the following 
steps were iterated until all diagnostics converged. 

First the \Halpha\ profile and an assumed \logg\ were used to estimate \teff. 
\feh\ was adjusted until weak \fei\ lines in the 5100\angs\ -- 5400\angs\ region 
were well matched; their strengths are largely gravity-independent.  The 
strength of weak \tii, \mgi, and \sii\ lines determined \alfe. Microturbulent 
velocity \vt\ was set by matching weak and strong iron and titanium lines. 
Gravity \logg\ was then checked by demanding matches for \mgi, \mgj, and the 
wings of the \mgb\ lines, for \fei\ and \fej, for \tii\ and \tij, and for \sii\ 
and \sij. This also constrains \teff, for the \mgj\ and \sij\ doublets arise 
from highly excited lower levels ($>$8\,eV), while the optical \tij\ and \fej\ 
lines originate from much lower levels ($<$3\,eV). Roughly two dozen \fei\ 
lines, a dozen \tii\ lines, and a dozen total \tij\ and \fej\ lines were 
detected for the most metal-poor stars, with double this number typical of the 
more metal-rich stars. The \mgj\ and \sij\ doublets at 4481\angs\ and at 6347.1 
and 6371.3\angs\ were seen in all cases, but \sij\ only marginally so in \hd 
84937 and \hd 19445.

Once the diagnostics began to converge, iterative adjustments were made to the 
Kurucz mid-UV atomic-line parameters. First, gf-values were modified for atomic 
lines identified in the laboratory. 
The relative contribution of identified lines to a blend was 
ascertained from the behavior of the blend in the hot star \hd 128167 versus the 
cooler ones, and in 
the higher-gravity stars such as \hd 19445 versus the lower-gravity ones. 
Damping constants were changed in some cases to match line profiles. Damping 
constants rather than gf-values were changed for the strong resonance lines 
well-observed in the interstellar medium for which gf-values are tabulated by 
\citet{mor91}. 

Next, for lines lacking identification, the 
laboratory line list was searched for transitions close enough in wavelength to 
perhaps be responsible. The gf-value of a nearby 
identified line was increased by up to 4.5 in the log, and the calculation 
repeated. This line was 
dropped if still not strong enough; otherwise its gf-value was adjusted to 
match. The high resolution 
of the \acena\ and solar spectra showed that most such identifications were very 
well matched in 
wavelength, and thus likely to be correct, when based on increases of $\leq$ 2.5 
in the log. 

For the two most metal-poor stars, these steps were sufficient to enable the 
continuum to be 
defined unambiguously throughout most of the mid-UV region. Its definition in 
more metal-rich stars 
was hampered by line crowding, yielding a pseudocontinuum that the modeling did 
not properly reproduce. 

Much of this pseudocontinuum mismatch appeared to be due to individual 
transitions missing from the line list. This was indicated by the very common 
occurrence throughout the mid-UV of unmodeled absorption features found at the 
same wavelengths in all stars, features whose strength steadily increased in 
going from weak-lined to strong-lined spectra. Most of these transitions are 
weak, consistent with the fact that the intrinsically strongest transitions of 
each species are the ones most likely to have been detected in the laboratory. 

Except just redward of the \mgi\ edge at 2512\angs, most missing lines are 
probably due to iron. Its abundance is high and the spectra are rich in lines of 
\fei\ and \fej, which are well-populated species in solar-temperature stars. Of 
the elements with similar first ionization potentials, only magnesium and 
silicon have comparable abundances; the mid-UV spectra of their low-ionization 
stages are generally much less complex, as are those of the more abundant CNO 
elements. 
While iron-peak elements such as Cr, Mn, and Ni have many transitions throughout 
the mid-UV, iron dominates because it is twenty or more times as abundant. In 
most halo stars above one-tenth solar metallicity, the relative abundances of 
these elements with respect to iron are within 0.2 dex of those of the Sun 
\citep[although larger differences may occur, especially at very low 
metallicity:][] {mcw95,nis00,pro00}. The mismatch that will result from 
assigning missing lines of these elements to iron should be mild. 

In the region immediately redward of the bound-free \mgi\ opacity edge at 
2512\angs, the probability is significant that missing lines may be due to 
magnesium rather than iron. Redward of the edge there are a very large number of 
transitions to known highly excited levels of the \mgi\ atom, and presumably a 
very large number of similar transitions to closely-spaced highly excited levels 
that are as yet unidentified in the laboratory. If many weak transitions redward 
of 2512\angs\ are due to magnesium but have been assigned to iron instead, our 
calculations will underestimate the absorption in this region in metal-rich 
alpha-enhanced populations. We see below that this region is poorly modeled in 
any case once metallicity exceeds one-tenth solar.

The modeling then proceeded iteratively by assigning tentative identifications 
and transition probabilities to each missing feature, then recalculating until 
an acceptable match was achieved or deemed unattainable with existing data (as 
for the regions just redward of 2550\angs\ and 2640\angs\ in the solar 
spectrum). Features were attributed to either low-excitation \fei, 
moderate-excitation \fei, or \fej, whose line strengths depend differently on stellar 
temperature and gravity (pressure) because of the Boltzmann and Saha equations. 
With a \teff\ $\sim$1000\kel\ higher than those of other moderately weak-lined 
stars, \hd 128167 proved exceptionally useful in making these assignments. The 
higher resolution of the spectra of the strong-lined stars \hd 128620 = \acena\ 
and the Sun was essential in discerning the wavelengths of the many features 
emerging above one-third solar metallicity. 

A brief examination was made of the gf-value changes for low-excitation \fei\ 
lines, namely those identified in the laboratory with (air) wavelengths falling 
between 2600\angs\ and 3000\angs, and with a lower level of 6000 -- 10000\cw\ 
and an upper level of 40000 -- 50000\cw.  Sixty-five such \fei\ lines have gf-
values from the laboratory measurements of \citet{obr91}; 52 were not changed, 
while 13 were changed by from $-0.6$ to $+0.4$ dex. 
Nine lines have gf-values from the laboratory measurements of \citet{fuh88}; 
five were unchanged, and four changed by from $-0.7$ to $-0.3$\,dex. Thus the 
mid-UV gf-values found in this work generally concur with the mid-UV laboratory 
measurements. 

In contrast, fifteen lines have theoretical gf-values from \citet{kur94}; four 
were not changed, 
but 11 were changed by from $-2.2$ to $+0.6$\,dex, all but one decreased. 
The theoretical gf-values of \citet{kur94} are seen to be subject occasionally 
to very large errors, usually overestimates. This can be attributed to the 
finite size of the Cray computer used in those calculations, which dictated that 
the number of levels included be truncated, and so upper levels were 
preferentially omitted from the model atom. Because the sum of the gf-values of 
transitions arising from a given lower state is normalized to unity, gf-values 
for the transitions included in the calculation will be overestimated because of 
the exclusion of others arising from the same level. Thus the calculations will 
overestimate the gf-value by increasing amounts as the fraction decreases of the 
actual possible upper states that are included in the atomic model. This 
fraction decreases to the blue, because higher upper levels are involved for a 
given lower level. Indeed, the severity of overestimates appears to increase to 
the blue: all the above changes made to the \citet{kur94} gf-values were $-1.5$ 
or lower below 2650\angs, but redward of this only one change this negative was 
required.  The overestimates among the \citet{kur94} \fei\ gf-value calculations 
are thus seen to be due to incompleteness and not to more fundamental 
difficulties. Undoubtedly many of the weaker mid-UV transitions are missing 
entirely from the predicted list.

For the hydrides, separate treatment was required, as the lines are naturally 
associated into bands, and the Kurucz hydride calculations are now several 
decades old. For OH, the theoretical calculations of \citet{gil01} (communicated 
in August 2000 by M. Bessell) were used wherever possible. For CH, the LIFBASE 
data \citep[e.g.,][]{luq99} were used (communicated in August 2000 by M. Bessell 
with lines cross-identified). 
Since these cover only the strongest bands, calculations must still rely on the 
Kurucz values for the others, notably the 4-3 band of OH with many weak but 
discernible lines in the Sun. A comparison of the Kurucz OH transition 
probabilities against those of \citet{gil01} showed strong trends, with 
discrepancies increasing at high rotation number J. To place them on 
approximately the same scale, the Kurucz transition probabilities for the OH and 
CH lines were decreased by an amount $\delta$ = $-0.25 -0.05*J_l$\,dex. 
Once missing atomic lines were largely filled in, a comparison was feasible of 
calculated OH and CH line strengths against those of the Sun and stars. Because 
solar OH lines were otherwise modeled too strong, all the transition 
probabilities were reduced by an additional 0.15 dex uniformly. In part because 
weak OH lines in the Sun were reasonably well modeled but strong ones were still 
too strong, a chromospheric mitigation of the surface temperature drop was 
invoked as described shortly. As illustrated below in Figs.\ 3$d$ and $e$, these 
changes suffice to reproduce both strong and weak 0-0 OH features in stars as 
well as the Sun. 

The calculations were first run using \citet{kur95} models, but the very metal-
poor stars showed a \teff\ deduced from \Halpha\ as much as 200\kel\ higher than 
that from fitting the mid-UV. That discrepancy was removed by adopting 
\citet{cas97} models, models in which convective overshoot has been turned off. 
We did not use the alpha-enhanced \citet{cas97} models, as they are available 
for only a single value of the alpha-element abundance enhancement.

Like the \citet{kur95} models, the \citet{cas97} models we used were calculated 
with \vt\ = 2\kms\ and a logarithmic solar iron abundance of $-4.37$ with 
respect to the abundance by number of hydrogen plus helium, which amounts to 
$-4.33$ with respect to hydrogen alone. 
These choices are about 1\kms\ higher in \vt\ than the $\sim$1\kms\ typically 
found for the solar flux spectrum, and about 0.15\,dex higher in iron abundance 
than indicated by the meteoric iron abundance and by several recent 
determinations for the Sun \citep[e.g.,][]{bie91,hol95}. 
Thus both sets of models were calculated with higher line blanketing 
than expected for a solar-type star of the nominal model metallicity, since iron 
lines dominate the UV and optical spectrum, and the higher \vt\ increases
the strengths of 
moderately strong lines of every species.

For each star including the Sun, a \citet{cas97} model was interpolated to the 
nearest 50\kel\ in \teff\ (25\kel\ for the Sun), 0.1\,dex in \logg, and 
0.05\,dex in \feh. The microturbulent velocity \vt\ was changed at all depths to 
the value indicated for each star in the table and figures below. To minimize 
the overblanketing effect, our models were interpolated to a nominal grid 
abundance 0.10 -- 0.15\,dex lower than that desired. To set abundances, the 
abundance of the interpolated model was then raised to the desired value and the 
relative abundance of iron dropped by 0.15 dex. The specific light-element 
abundance elevations required by the optical observation were adopted, or 
assumed to be zero for the Sun and \acena, and all species equilibria were 
recalculated. 
This procedure incorporates the specific \vt\ and iron and light-element
abundances set for the model in calculating both the ionization and molecular 
equilibria and the line and continuous opacities for its spectrum. Concerning 
the model structure, the procedure compensates for the effects of overblanketing 
due to the high choice of iron abundance, but not for the light-element ratio 
nor for the microturbulence.
We anticipate very mild errors in model colors and temperature structure until 
spectra become dominated by molecular lines or otherwise are considerably 
stronger-lined than any of these. The comparison of model and observed colors 
(\S 6) bears out this expectation.

For the Sun, a model with \teff\ = 5775\kel, \logg\ = 4.4, \feh\ = 0, \alfe\ = 
0, and \vt\ = 1.0\kms, was interpolated in the above way.
For \hd 128620 = \acena\ we adopted a similarly interpolated model with \teff\ = 
5800\kel, \logg\ = 4.3, \feh\ = +0.15\,dex, \alfe\ = 0, \vt\ = 1.0\kms. While 
\citet{neu97} found \teff\ = 5830\kel\ and \feh\ = 0.25\,dex, most other 
analyses have derived somewhat lower \teff\ and \feh\ values. At \teff\ = 
5800\kel\ and \vt\ = 1.0 \kms, \feh\ = 0.15\,dex provides a better match to the 
mid-UV spectrum of \acena. 

Even with the use of \citet{cas97} models interpolated as above, another major 
type of mismatch persisted in all strong-lined stars: cores of lines formed near 
the surface appeared too strong in the calculations, especially in extremely 
strong mid-UV lines such as those of \mgj\ and \mgi\ or \fej\ and \fei. This is 
definitely not due to wayward gf-values or damping constants, because of the 
trend seen in the mismatch. Profiles calculated for the strongest lines in the 
most metal-poor spectra generally matched or were somewhat too high in the core, 
but were somewhat too low in the core in stars of intermediate line strength, 
and became too low across the the central ten \AA ngstroms or more in the 
strongest-lined stars. (See Fig.\ 3$a$ below for residual trends of this 
nature.) As noted above, a related mismatch was seen within the OH spectrum 
itself: in any given strong-lined spectrum, notably that of the Sun for which 
the model parameters are extremely well known, weak OH lines were matched at a 
higher oxygen abundance than strong lines. The possibility discussed in \S 5 
that the solar continuum may be drawn too low would lead to a mismatch of 
opposite sign. To the extent that the gf-values are reliable, then, this 
indicates an error in the temperature structure of the model. 

The most obvious possibility is a chromosphere, which is not incorporated in 
flux-constant, static models such as those of SYNTHE. The solar chromosphere, 
for example, is seen in images of the limb and evident from emission reversals 
in the cores of strong spectral lines such as \mgj. It is generally modeled by 
an increase in temperature in relatively shallow layers of the atmosphere 
\citep[e.g.,][]{fon99}. Any of several non-radiative heating sources extending 
from the photosphere through the chromosphere into the corona could be 
responsible \citep[e.g.,][]{sch95,sch99,ste00}.

We consequently changed the temperature of individual points in each 
interpolated stellar model with \feh\ $\gapprox -1.0$, beginning just below the 
point where the model becomes optically thin at 5000\angs\ (typically at depth 
number 51 or 52). (We tried this for the more metal-poor models as well, but the 
profiles of all the strong lines became much too high near the core, so we left 
these models as they were.) To best fit both optical and mid-UV spectra 
simultaneously, we increased the temperature at model depths 28 through 48 by 
10, 20, 30, 40, 50, 60, 70, 80, 90, 100, 110, 120, 130, 140, 150, 150, 140, 120, 
90, 60, and 30\kel\ respectively. This temperature enhancement may be considered 
an approximate chromosphere, though it is based entirely on empirical fits of 
one-dimensional spectral line calculations to observed spectral line profiles. 
It incorporates neither energy transport considerations nor matches to spatially 
resolved structures, both of which lie beyond the scope of this work. 

\section{Matching Mid-UV Spectral Calculations}

The degree to which our mid-UV and optical spectral calculations match spectra 
of real stars is illustrated in the figures. The first three show mid-UV 
spectra, and the last, optical. Within a figure, panels are ordered by 
wavelength. Flux (central intensity for the solar mid-UV spectrum) is plotted 
versus wavelength in air. The spectrum calculated for each star (light line) is 
plotted on top of that observed (heavy line).

In Figure 1,  the global mid-UV fit is shown for 
the nine stars with absolute fluxes. All spectra have been Gaussian 
smoothed to a FWHM resolution of 500 (550 for \hd 128127 and \acena). 
Each spectrum is displaced vertically by $\sim$20\% from the one below. For each 
star a single mid-UV 
continuum normalization constant was adopted, as listed on the left. On the 
right, the HD number of 
each star appears in bold. Below are found the model-atmosphere parameters 
adopted for the 
corresponding calculation: \teff\ (in \kel), C wherever the chromosphere was 
modified as described at the end of \S 4, log $g$, \feh, \vt\ (in \kms). On the 
following line appear the values in dex of the light-element 
abundance enhancements [Element/Fe] (in dex), first that of oxygen, then those 
for Mg, Si, Ca, and Ti. 
(The latter four are grouped together wherever they are the same.)
While the fits are generally good, serious discrepancies are noted in \acena\ at 
the bottom of the figure near 2550 -- 2600\angs\ and 2640 -- 2710\angs, regions 
of relatively little line absorption in the more metal-poor spectra.

Figure 2 illustrates the effect on mid-UV fits when model parameters are changed 
somewhat. Two of the 
stars of Fig.\ 1 are depicted: the moderately weak-lined \hd 94028 (top two 
spectra) and the rather strong-lined \hd 184499 (bottom three). Star and model 
identifications and normalization constants are indicated as in Figure 1. The 
lower plot for each star represents its best fit, 
a replot of that star's comparison in Figure 1. 

Comparing the bottom plot to the middle plot shows the effect for \hd 184499 of 
decreasing \mgfe\ by 0.3\,dex to the solar \mgfe\ ratio. 
Although minimized by the adoption of a normalization constant 11\% higher, 
the mismatch is still evident. 
The strengths of the wings of the strong \mgj\ doublet at 2800\angs\ (and of 
\mgi\ at 2852\angs) are poorly reproduced, as is the overall pseudocontinuum 
level near and below the bound-free edge of \mgi\ at 2512\angs. Similar 
calculations run for \hd 94028 show a more dramatic effect in the \mgj\ and 
\mgi\ line wings, which are weak enough that little renormalization is possible, 
but a smaller effect near 2512\angs, where the H$^-$ opacity is dominant because 
of the reduced 
magnesium abundance.

Comparing the plot just above the best fit for each star to the best fit itself 
shows the degeneracy of the mid-UV spectra with respect to a simultaneous change 
in \teff, \logg\ and \feh\ designed to preserve overall line strength and 
ionization equilibrium. For each plot just above the best fit, 
\teff\ was decreased by 100\kel, \feh\ and \logg\ both concurrently decreased by 
0.1\,dex, and 
the spectrum renormalized accordingly. 
This change is very difficult to detect from fitting the mid-UV slope alone at 
these metallicities and this resolution. The increased metallicity reduces the 
blue 
pseudocontinuum relative to the red at the same time that the increased \teff\ 
raises it. The change is 
marginally detectable in \hd 94028 from the mismatch of low-opacity windows, but 
not in \hd 184499 because the overall match is less satisfactory. Such a change 
does lead to a reduction in the normalization parameter, by 11\% for \hd 94028 
and 17\% for \hd 184499. In principle this might be detected by other means, but 
only if \mgfe\ is properly modeled. For as seen above, taking \mgfe\ = 0 (as in 
the Kurucz flux distributions) can largely compensate in stars such as \hd 
184499 that show a light-element enhancement. 

Our conclusion, then, is that if the mid-UV normalization constant is treated as 
a completely free parameter (as here), the mid-UV flux distribution by itself 
can currently be used to set \teff\ only in MSTO stars with \feh\ $\leq -1.5$. 
While the mid-UV was useful in demonstrating the need for models with no 
convective overshoot and with an approximate chromosphere, our determinations of 
\teff\ for the more metal-rich MSTO stars have relied on the optical spectra 
instead.
We show in Fig.\ 4 below that the Balmer line profiles and the strengths of 
other high-excitation lines are well matched in the optical spectra of all stars 
at the same \teff\ as the best-fit mid-UV spectrum, and in \S 5 that model 
colors match observations as well, so that \teff\ is indeed very well 
established here. However, our goal of \teff\ determination for turnoff stars of 
all metallicities from mid-UV spectra alone is not yet met. We are not able to 
calculate the low-opacity windows blueward of 2900\angs\ accurately enough at 
near-solar metallicity, as we now demonstrate.

Figure 3 plots at higher resolution and higher scale the same mid-UV spectral 
comparisons as are shown in Figure 1, plus that of the Sun. 
Panel $a$ shows 2627.5 -- 2630.5\,\AA, dominated by strong lines. The other 
panels show regions of locally low line blanketing: $b$, 2642.5 -- 2645.5\angs; 
$c$, 2900.0 -- 2903.0\angs; $d$, 3082.5 -- 3085.5\angs; $e$, 3100.0 -- 
3103.0\angs.  In all plots in Fig.\ 3, the spectrum of \hd 184499 is offset by 
30\% from that of \hd 128620 = \acena, which in turn is offset by 50\% from that 
of the Sun, for which 1.0 marks the height of the true continuum. The same 
normalization constant as in Fig.\ 1 has been adopted for each stellar spectrum, 
as was a normalization for the Sun 1.10 times that given in the \citet{kur81} 
atlas, as noted on the left. Identifications are provided at the top for the 
strongest lines as calculated in the solar spectrum. Manually added lines are 
flagged with a colon following the decimal digits of the wavelength in \AA 
ngstroms. Next appears the species, the lower excitation potential in eV of 
atomic transitions or the band and line of molecular ones, the digits following 
the decimal of the residual flux in the unbroadened line of the solar spectrum 
calculation, and the log gf-value adopted.

In Fig.\ 3$e$ of the 3100.0 -- 3103.0\angs\ region, calculated and 
observed spectra agree reasonably well in all stars, including \acena\ and the 
Sun. 
No ``missing'' lines are strong enough to merit explicit identification at the 
top of the figure.  True continuum is reached in each spectrum, but fleetingly, 
near 3102.5\angs; its placement is sensitive to the treatment of adjacent weak 
lines. An unblended 0-0 OH line appears at 3101.230\angs, next to a near-
continuum high point; both are modeled well in all spectra. The strong atomic 
lines are well reproduced, as are two moderately strong CH lines at 3100.120 and 
3100.170\angs. 

Fig.\ 3$d$ offers a further picture of how well both continuum and OH features 
are modeled. Another continuum window appears near 3084.7\angs, but again with a 
weak line superimposed in the solar and \acena\ spectra. The relatively 
unblended OH 0-0 band line at 3084.896\angs, used in several oxygen-abundance 
investigations, is well reproduced over the factor-of-100 metallicity range of 
the stars in whose spectra it is detected, including the Sun itself. Nearby at 
3085.2\angs\ is a blend dominated by a stronger 0-0 OH line; it too is 
reasonably well matched in all stars. Recall that to achieve this agreement, the 
\citet{gil01} theoretical gf-values for OH were reduced by $0.15$\,dex; 
otherwise these features and that of Fig.\ 2$e$, plus the overwhelming majority 
of other OH lines, would be too strong in the Sun. That is, unless the solar 
continuum were raised. 

The solar continuum appears to be well defined, and the OH lines of all bands 
matched to $\sim$0.1\,dex, throughout the 3000 -- 3100\angs\ region. Because 
there are so few continuum windows, however, we cannot rule out a systematic 
underestimate of the continuum in \acena\ and the Sun of perhaps 5\% or 10\%. 
Roughly speaking, raising the \acena\ mid-UV continuum would allow it to be 
modeled at an \feh\ value more consistent with that of \citet{neu97}, while 
raising the solar continuum might eliminate the present need to reduce the 
theoretical OH gf-values by 0.15\,dex. As noted at the end of \S 6, it would 
also bring the mid-UV normalization constants of the Sun and \acena\ closer to 
their expected values. 
A definite answer could be obtained by deriving the stellar parameters of 
\acena\ from an optical spectrum in the same way as for the other stars, and 
from modeling an improved spectrum of the solar near-UV.

Random continuum variations approaching 5\% occur for two, and possibly all 
three, of the stars that were noted in \S 3 as observed through a smaller 
aperture. For the continuum seems appropriately placed for all stars in Fig.\ 
3$e$, but in Fig.\ 3$d$ the continuum of the \hd 106516 observed spectrum 
appears too low, and in Fig.\ 3$c$  that of \hd 128167 may also be too low. Such 
variations might also be present in the \hd 128620 = \acena\ spectrum as well, 
but remain undetected because of the difficulty of matching its multitude of 
lines. 

Fig.\ $3c$ shows another low-opacity window at 2900\angs. Here the number of 
missing lines added remains small, and agreement is nearly as good. OH lines are 
less important, being generally minor constituents of blends. However, windows 
of true continuum have disappeared from the solar and \acena\ spectra, and are 
fleeting at best in the echelle spectra of lower resolution that were acquired 
for mildly metal-poor stars. Continuum placement depends not so much on OH 
features but rather on whether all missing lines have been spotted. Clearly that 
is not the case at solar metallicity, where weak missing lines apparently 
overlap to the extent that they cannot be distinguished as individuals. Although 
the cumulative effect of their neglect on the total absorption is small, is it 
significant for the continuum placement, which consequently cannot be determined 
as well as was the case above from fits such as this. It is placed here and for 
the remaining panels by adopting the same normalization factors required to 
match Figs.\ 3$d$ and $e$.

By 2645\angs, in 
Fig.\ 3$b$, the fits have deteriorated considerably. This is especially true at 
and above solar metallicity, where the numbers of missing and identified lines 
noted at the top are comparable.
True continuum still appears in the weakest-lined spectra near the top of the 
panel, but elsewhere it is gone, lost to a forest of weak lines. Although the 
fit to the metal-poor stellar spectra is still reasonable, the high-resolution 
spectra of the \acena\ and the Sun reveal that too few missing lines have been 
added, often too strong and in the wrong places. For the adjacent region from 
2645 to 2700\angs, the sheer number of missing lines precluded their reliable 
identification at the low resolution of the low-metallicity stellar spectra.  
Because many were simply omitted, the calculations overestimate the spectral 
flux in that region -- mildly in moderately metal-poor stars but seriously at and 
above solar metallicity, according to Fig.\ 1.

 From Fig.\ 3$a$, it is clear that these difficulties are reduced in regions 
dominated by strong atomic lines. There the fit is dictated more by the ability 
of the calculations to match the cores and the damping wings of strong lines, 
and not by modeling the weak or missing features. These are largely suppressed: 
although the region shown is blueward of that of panel $b$, only one missing 
line is noted in it. The wings of the strong lines are reproduced well in the 
four top spectra but begin to be somewhat too broad near the core in \hd 106516 
and \hd 201891. The entire $\pm$1\angs\ about the core is too deep in \acena\ 
and the Sun. This could perhaps be remedied by continuing the temperature 
enhancement to shallower surface layers than those indicated in \S 4. 

Figure 4 shows fits to the optical spectra of the Sun and all stars except 
\acena. 
Identifications are given for the strongest lines in the solar spectrum. 
Since the optical spectra were not fluxed, the continuum was normalized to 
unity. 
The same models were used to calculate the synthetic spectra as in Figures 1 and 
2.
The wavelength regions in panels $a$ -- $e$ depict a variety of magnesium lines 
and those of other heavier elements, to illustrate the extent to which the 
spectral calculations match features of a wide range of excitation, ionization, 
and strength. 

Figure 4$f$ shows \Halpha, to illustrate the satisfactory choice of \teff. 
The normalization of the \Halpha\ profile was aided by its 
position near the center of an echelle order, and was done by interpolating the 
normalization 
functions of the two adjacent orders, centered $\pm$75\angs\ away. This was 
straightforward, for the \Halpha\ line is much narrower than this and the 
continuum is very well defined by a spline fit of low order; moreover, the 
normalization functions for the adjacent orders were very similar. 
Note that most absorption lines in this panel are telluric, not stellar, as 
evidenced by their shift in apparent wavelength from one star to the next.

Included in 
Fig.\ 4$f$ above the best fit for \hd 94028 is the spectrum for the cooler \hd 
94028 model whose mid-UV spectrum is shown in Fig.\ 2. This illustrates how 
the fit to \Halpha\ deteriorates when models of 100\kel\ lower \teff\ are 
adopted. The \Halpha\ profile 
is relatively insensitive to gravity in this case: calculations adopting a model 
that differed
from the best-fit model only in having \logg\ 0.2\,dex larger produced a 
profile that deviated only near 6560\angs\ and 6565\angs, 
by about the thickness of the heavy line. Similar calculations showed no 
sensitivity of the \Halpha\ profile to the chromospheric elevation of 
temperature, nor to changes of 0.3\,dex in iron abundance or light-element 
ratio. However, \citet{kur95} models required a \teff\ 200\kel\ hotter, thanks 
to their inclusion of convective overshoot.

Panels $a$ -- $e$ of Fig.\ 4 show that all the optical magnesium features are 
well matched using the same magnesium abundance and the same model as for the 
best fit to the mid-UV and to \Halpha. Panel $a$ plots the region near the very 
high-excitation \mgj\ doublet at 4481.2\,\AA; $b$, the ground-state \mgi\ line 
at 4571.1\angs; $c$ and $d$, the \mgb\ lines at 5167.3, 5172.7, and 5183.6\angs; 
and $e$, 
the high-excitation \mgi\ line at 5711.1\angs. These panels also demonstrate 
that the calculations match strengths and profiles of other atomic lines 
regardless of species, excitation, or strength. Among the lines depicted are 
\tii\ at 5173.7\,\AA, \tij\ at 5185.9\,\AA, ground-state \fei\ at 5166.3\angs, 
\fej\ at 4576.3\angs, and \sii\ at 5708.4\angs. 
These panels also contain missing features: not modeled at all in the 
calculations are the absorption lines seen near 4480.8\,\AA, 5170.7\angs, and 
5181.3\angs.  They are not telluric, since they appear at the same position in 
the stellar spectra, and are especially visible in the two stronger-lined stars. 

\section{Results and Comparisons} 

Our results for the parameters of the eight metal-poor turnoff stars analyzed 
here are listed in columns 12 -- 15 of Table 1. The \teff\ values found for 
these stars by works cited in \S 1 are included for reference. 
Our \teff\ values are seen to fall in the middle of the range of previous 
results, preferring neither the hotter nor the cooler values. Unfortunately, the 
number of stars is too small and the scatter too large for more definitive 
conclusions.

The reliability of our \teff\ values is supported by a comparison of model 
colors to those observed. 
For each star including the Sun, we interpolated the $UBV$ and \bmy\ colors and 
$V$ magnitudes calculated by \citet{cas97} to the \teff, \logg, and \feh\ values 
of each model we used. We have made no reddening corrections, nor any 
corrections for the temperature enhancement invoked in the more metal-rich 
models as described at the end of \S 4. The ninth, tenth, and eleventh columns 
of Table 1 show the difference in \bmv, \umb, and \bmy\ between the observed and 
calculated colors for each star. The agreement is extremely good in all colors 
over the entire metallicity range. The mean difference is $-$0.010, $-$0.018, 
and 0.000\,mag in \bmv, \umb, and \bmy\ respectively. The standard deviation of 
an individual difference is 0.010, 0.026, and 0.015\,mag, marginally larger than 
typical observational errors alone. The color differences are small even for 
\acena, for which both metallicity and \teff\ are influential (we interpolated 
to a model \feh\ = +0.05), and for which we have no optical spectrum and so did 
not derive parameters. For the other stars, \bmv\ and \bmy\ are sensitive 
primarily to $T_{{\rm eff}}$; their deviations correspond to $\pm$45\kel\ with 
no allowance for observational error. Their observed colors are slightly bluer 
than those of models, but the size of the differences points to a scale error in 
\teff\  of less than 50\kel.

We thus estimate an uncertainty of $\pm$50\kel\ in each \teff\ determination, 
from this and the poorer quality evident in Figs.\ 2 and 4$f$ 
of the fits to mid-UV fluxes and \Halpha\ profiles as \teff\ is increased by 
100\kel, plus the deterioration of the fit shown in Fig.\ 4$a$ to \mgj. 
 From plots like those of Fig.\ 4 at other wavelengths,
we estimate the uncertainty in \feh\ to be $\pm$0.05 dex at fixed \teff, 
increasing to 
$\pm$0.1 dex when the \teff\ uncertainty is included. The mid-UV slopes and 
\Halpha\ profiles are only 
weakly sensitive to \logg, so our results for \logg\ are susceptible (as are 
those of other groups 
relying on ionization equilibria) to errors in gf-value scales and to possible 
non-LTE effects. 
We have not yet taken full advantage of the technique of matching damping wings 
of strong lines, incorporating improved treatments of hydrogen broadening such 
as those of \citet{ans95} and \citet{bar97}. It would be valuable to do so given 
the constraints this would offer on alternative measurements of the fundamental 
properties of these standard stars, for example parallaxes from HIPPARCOS and 
other space missions. 
However, the following comparisons of observed and model fluxes indicate that 
the \logg\ values are good to $\pm$0.1\,dex as they stand. 

We first compare the UV normalization constant with that expected. According to 
\citet{mih81}, the ratio of observed to emitted stellar flux is $f/F = 
\theta^2/4$, where the angular diameter $\theta$ = 2$R$/$D$, $R$ being the 
stellar radius and $D$ its distance. For each star we calculate $\logr$ in solar 
units as $0.5 (\log M - \logg + 4.4377)$, assuming masses $M$ = 0.8\msun\ for 
for the metal-poor stars based on stellar isochrones \citep{vdbet00} and 
1.15\msun\ for \hd 128167, and adopting 1.16\msun\ for \acena\ as derived from 
its orbit by \citet{pou99}. The distance in solar units is found from the 
parallax, and the theoretical flux normalization constant follows from $F/f$ 
times 4.2545 $\times 10^{10}$ to convert to square steradians. The ratio of this 
theoretical constant to that actually used to scale the observations in Figs.\ 1 
and 3 is listed in column 7 of Table 1. 

Agreement is good for all stars but one. For \acena, the ratio is 0.46, 
suggesting an observed flux that is too low. Indeed, \citet{lin96} saw this in 
their 1995 mid-UV \acena\ HST observations with GHRS: relative to earlier GHRS 
and IUE observations of \acena, ``our fluxes are lower (by factors of 5.6 and 
4.8 for the \mgj\ and \fej\ spectra, respectively), presumably because of the 
star's being misplaced at the edge of the small aperture.'' From a visual 
comparison with Fig.\ 1 of \citet{lin96}, the fluxes we are using for the 
\citet{lin00} STIS \acena\ spectrum should be raised by a factor of 2.1. This 
brings the \acena\ ratio to 0.96, and results in an 
average of the nine stellar ratios of 1.06 $\pm$ 0.05, with an individual 
standard deviation of 0.15. 
For the Sun, the ratio is 0.91 given the scale factor of 1.10 noted above, which 
is within the stated normalization uncertainty but still low. These \acena\ and 
solar ratios would be closer to unity were the continuum levels for their 
spectra drawn higher by 5\% -- 10\%, a possibility raised in \S 5. As matters 
stand, errors in continuous opacity are limited to $<$10\% nonetheless.

The apparent $V$ magnitude may also be compared to that expected from the model. 
From the $V$ magnitudes tabulated with the model colors, we calculated the 
difference in the stellar model $V$ with respect to the solar model, added this 
and $5 \log(D/R)$ to the observed solar $V$ = $-26.7$, then subtracted this from 
the observed stellar $V$ magnitude. Again, no reddening or other corrections 
were applied. The resulting differences are shown in column 8 of Table 1. The 
agreement is reasonable: the theoretical magnitudes average 0.09 $\pm$ 0.06\,mag 
brighter than those observed, with a $\pm$ 0.18\,mag individual standard 
deviation. 

\section{Summary and Discussion}

In brief, we have calculated high-resolution mid-UV spectra from first 
principles, after modifying line parameters and assigning line identifications 
for missing lines based on changes seen among spectra of stars spanning a range 
of well-determined values of $T_{{\rm eff}}$, log $g$, and \feh. The resulting 
calculations provide a good fit for turnoff stars of all metallicities redward 
of 2900\angs. Previous discrepancies in fitting the solar pseudocontinuum (noted 
in \S 1) are attributed not to missing continuous opacity, which is good to 
$<$10\% near 3100\angs, but rather to missing or miscalculated line opacity. 

To improve the accuracy of the mid-UV gf-values, damping constants, and line 
identifications derived here, we have determined temperatures, gravities, 
microturbulence, and abundances for eight stars by calculating fits to optical 
as well as UV spectra. For stars more metal-poor than one-thirtieth solar, the 
fit to the mid-UV spectrum was satisfactory using laboratory lines alone, and 
can by itself determine \teff. For the more metal-rich stars, lines missing from 
the laboratory list compromise the goodness of fit of the mid-UV spectrum. A 
100\kel\ increase in \teff\ is seen to be difficult to detect from mid-UV 
spectra alone, as long as \feh\ and \logg\ are increased by 0.1\,dex to 
compensate. This \teff -\feh\ degeneracy parallels the age-metallicity 
degeneracy in elliptical galaxies \citep{wor94,wor99}. As with galaxies, it is 
reduced here by considering the Balmer lines, with the advantage that stellar 
models need not include Balmer-line emission nor composite populations. Here 
\teff\ was set for stars with \feh\ $\geq -1.5$ by matching the \Halpha\ line 
profile and the strengths of other high-excitation features. 

For all the metal-poor stars, the excellent match between calculations and 
observations of both optical and mid-UV spectra, and between model and observed 
colors, indicates an uncertainty of $\pm$50\kel\ in an individual \teff\ 
determination, and a similar uncertainty in the \teff\ scale.
Moreover, the stellar $V$ magnitudes generally agree to 0.1\,mag, and the mid-UV 
normalization constants to 10\%, with those expected from the stellar angular 
diameter found from the observed parallax and the model \logg, assuming 
reasonable masses. 

In this process, it was found that these optical diagnostics and the slope of 
the mid-UV continuum could only be fit simultaneously when using models of 
\citet{cas97} in which convective overshoot has been turned off. Our work thus 
agrees with that of \citet{cas97}, who found that their models better reproduce 
the Balmer profiles in the Sun than do those of \citet{kur95}. 
Downloading the \citet{cas97} models from the Kurucz web site at 
http://cfaku5.harvard.edu is 
recommended for all work where the results depend upon the temperature structure 
in 
deep continuum-forming regions. 

Residual discrepancies in fitting the profiles of very strong atomic lines, and 
the strengths of strong versus weak solar OH lines in the 3000 -- 3100\angs\ 
region, are reduced here by mimicking a chromosphere towards the surface of 
every model with \feh\ $\gapprox$ $-1$. As described at the end of \S 4, the 
model temperature was raised in the region just above the depths where optical 
continuum and weak lines are formed. While this procedure was necessary to match 
strong-line cores and OH line strengths over a very broad range of metallicity 
with a single set of gf-values, it is definitely an oversimplification. Energy 
and stability considerations remain to be examined. Fits might be improved with 
a different choice of temperature enhancement. Indeed, a two-stream model might 
be found necessary. In the meantime, line cores also will be subject to 
uncertainties due to the presence of chromospheric emission. \citet{pet97} 
demonstrate that \mgj\ emission is always present in solar-temperature stars, 
with a strength that increases with increasing metallicity. Emission also 
increases with increasing stellar activity \citep{ayr95}. Since activity in turn 
is associated with rapid rotation, stronger \mgj\ emission appears commonly 
among young stars and close binaries. Consequently, no single quiescent 
chromosphere appears capable of modeling the very core of the profile of a 
strong line in all turnoff stars. 

The reliability of our \teff\ values and the good match achieved for OH lines 
enable a second look at the oxygen abundances of metal-poor stars, which are 
still uncertain as noted in \S 1. Our values for \teff\ are intermediate, 
suggesting that \teff\ is not the sole cause of the oxygen discrepancies. We 
defer a more detailed discussion to a future paper by \citet{pet01}, who compare 
oxygen abundances derived from the OH lines with those from the permitted and 
forbidden lines of atomic \oi\ in these stars. 

The manual addition of lines to the model line list, to match otherwise missing 
features seen in the observed spectrum, is clearly necessary for even an 
approximate representation of the mid-UV spectrum of solar-temperature stars of 
one-tenth solar metallicity and higher. Nonetheless the procedure has its 
important limitations. One is that as overall line strength increases, the 
proportion of missing lines goes up dramatically; disentangling them requires a 
broader range of spectra of high quality, as discussed below. Another is that 
the excitation or ionization of the transition could be misjudged; the 
ramifications are discussed in the next paragraph. A third is that many missing 
lines may not be due to iron as assumed, but to other elements. As discussed in 
\S 4, this is potentially serious redward of the bound-free absorption edge of 
\mgi\ near 2512\,\AA. Elsewhere the effect should be minimal, because although 
other iron-peak elements contribute significant line absorption throughout the 
mid-UV, the relative abundances of all other iron-peak elements with respect to 
iron change little from one late F or early G main-sequence star to another in 
both halo and disk populations in our Galaxy. To be sure, the list generated by 
this procedure cannot be expected to reproduce the spectrum of a peculiar A or F 
star, for example a Cr-enhanced star, in which the iron-peak ratio has been 
dramatically altered \citep{jas90}. 

More generally, the list is not suitable for modeling stars whose parameters lie 
well outside the range of those analyzed here. In such cases, any difference 
becomes important between the assigned and the actual excitation (and 
ionization) potential of the transition responsible for a line, leading to a 
significant misrepresentation of its strength. Consequently, the present list 
cannot be expected to reproduce the spectra of stars hotter than early F nor 
cooler than early G, nor giants, as well as stars of metallicity substantially 
higher than solar for which lines are still missing. The list could be made 
suitable for those stellar types by extending the present analysis to include 
such stars, provided high-resolution, high S/N mid-UV spectra are available in 
which mismatches induced by errors in the list can be visually identified and 
corrected. We discuss this further below. 

Conversely, the list as it stands should do as well as indicated by Fig.\ 1 and 
Fig.\ 3 in reproducing the mid-UV spectra of any other late F or early G star of 
solar metallicity or less, provided a suitable model atmosphere is chosen. This 
holds because once the gf-value and damping constant of a line transition are 
set, elemental abundance and the Boltzmann and Saha equilibria at each point in 
the atmosphere fully determine line strength in quiescent stars such as these. 
Dynamical effects can be ignored, and non-equilibrium effects (if any) are 
slowly-varying functions of spectral type, hence absorbed into the gf-value 
determinations. The only exception is the chromospheric emission noted above. 
Were the line list improved to the point that a reliable match is achieved to 
the mid-UV spectra of F and G stars of solar metallicity and above, we could 
expect to use the mid-UV slope to at least constrain age and metallicity, 
regardless of its overall level or the appropriate light-element abundance 
ratio, as these could be incorporated in a grid of calculations covering the 
entire relevant parameter space.

Spectral indexes offer an alternative means by which mid-UV spectral 
calculations alone might constrain metallicity and \teff\ of a star or age of a 
stellar system. The principle is to construct indexes based on spectral regions 
dominated by features whose behavior breaks the age/metallicity degeneracy and 
elucidates the alpha enhancement, examples of which are the Balmer lines and 
ratios of \fei/\fej\ and \mgi/\mgj\ line strengths. 
Spectral indexes have indeed been used as diagnostics of age and metallicity in 
ellipticals by several of the groups mentioned in \S 1. 
For example, \citet{lot00} have compared mid-UV and optical line indexes they 
calculated from Kurucz flux 
distributions against mean values observed for the stars considered by 
\citet{fan90}. Unfortunately, 
poor agreement was seen for all but two mid-UV indexes, the slope of the 
pseudocontinuum between 2600\angs\ and 3100\angs, and blended absorption near 
2538\angs. 

We should be able to improve upon some of those index calculations immediately. 
As Figs.\ 1 and 3 illustrate, the current spectral calculations should be 
reasonably reliable at all metallicities in spectral regions redward of 
2900\angs, and for those dominated by strong lines blueward of that. This 
includes several indexes already in use, notably the \mgj\ and \mgi\ lines at 
2800\angs\ and 2852\angs\ and line blends in the 3000+\angs\ region. We plan to 
begin such index calculations shortly, and make them generally available.

Blueward of 2900\angs, however, our calculations are not yet reliable enough. 
Fig.\ 1 shows that our calculated spectra of solar-metallicity stars 
substantially overestimate the true fluxes in those wavelength regions not 
dominated by strong lines. Somewhat ironically, our calculations thus suggest 
that neither 
of well-behaved \citet{lot00} indexes is currently reliable at near-solar 
metallicity. 

The problem undoubtedly is lines still missing from the line list. As seen in 
Fig.\ 3,
missing lines often overlap in the solar spectrum to such an extent that they 
could not be individually discerned, and so their species and wavelength 
assignments became arbitrary. This was true especially in the 2550 -- 2600\angs\ 
and the 2640 -- 2710\angs\ regions, where we did not complete the process. This 
is the reason for the excess flux seen in the calculations at solar metallicity 
in these regions. 

This could be remedied were spectra obtained of the same quality as that of 
\acena\ for stars of the line strength of \hd 184499 that span a range of 
properties, notably ${T_{{\rm eff}}}$. For example, the spectra of \hd 184499 
and a Hyades F star with \teff\ $\sim$6500\kel\ should have about the same 
overall line strength: the Hyades star's higher abundance, \feh\ $\sim$+0.15, is 
offset by its higher $T_{{\rm eff}}$ at the rate of 0.1\,dex per 100\kel, as 
Fig.\ 2 illustrates. Because line crowding is much lower than in \acena, 
comparing high-resolution mid-UV spectra of \hd 184499 and a sharp-lined Hyades 
F star could show where the strongest lines are that are missing at solar 
metallicity and suggest their identities from the changes observed with $T_{{\rm 
eff}}$. The \acena\ and solar spectra could then be used to fill in the rest. 
Analyzing an additional star closer in \teff\ to \hd 184499 but with \alfe\ = 0 
would show explicitly which transitions are due to magnesium in the 2550\angs\ 
region, and whether OH has been properly characterized throughout the 2900 -- 
3100\angs\ domain.

At that point, the full arsenal of indexes representing the strengths of mid-UV 
features could be calculated from the spectra, and their behavior with \teff, 
\logg, \feh, and \alfe\ assessed explicitly. This is a major advantage of 
calculations from first principles over empirical libraries, which are confined 
to the spectra of stars available nearby. It is a major goal of this project. 
Indexes should result with significant diagnostic power for the integrated-light 
spectrum of an old stellar system.

\acknowledgements 
We are indebted to S. Allen of U.\ C.\ Santa Cruz 
for writing the initial scripts to convert the VAX version of Kurucz codes
into UNIX versions. We are grateful to M. Bessell of Mount Stromlo and Siding 
Spring Observatories 
for communicating the \citet{gil01} and LIFBASE gf-value calculations and 
programs for their comparison against the Kurucz values, 
to J. Valenti of the Space Telescope Science Institute for providing the \hd 
128620 spectral reductions, 
to T. Misch of Lick Observatory for obtaining the blue spectrum of \hd 114762,
and to J. Fulbright of the Dominion Astrophysical Observatory for contributing 
his optical echelle spectra. We thank Fulbright and R. P. Kraft of Lick 
Observatory, 
R. L. Kurucz of the Harvard-Smithsonian Center for Astrophysics, 
T. M. Lanz and S. R. Heap of the NASA/Goddard Space Flight Center, 
D. C. Morton of the Herzberg Institute for Astrophysics, 
and R. W. O'Connell of the University of Virginia 
for useful discussions. \acksim. Support for this work was provided by 
NASA Astrophysics Data Program contract S-92512-Z, 
STScI grants GO-07395, GO-07402, and AR-8371, and NSF grants AST 99-00582 and 
AST-0098725, to Astrophysical Advances, 
and by NASA ADP grant NAG 5-7104, NASA LTSA grant NAG 5-6403, and STScI grant 
GO-06607 to the University of Virginia.

\clearpage


\figcaption[fig1.eps]{\label{hd9glo.240}} Plots are shown comparing observed 
(heavy line) and 
calculated (light line) spectra for the nine stars with fluxed mid-UV spectra 
included in this study. To the 
right, the name of each star in bold is accompanied by the model parameters used 
for the corresponding 
calculation.  At the left is the mid-UV normalization constant for each 
comparison.

\figcaption[fig2.ps]{\label{hd9tglo.240}} Plots as in Fig.\ 1 are shown 
comparing the fits for two 
stars with small changes in model parameters.

\figcaption[fig3a-e.eps]{\label{hd0grnog}} The same as Fig.\ 1, but on a larger 
scale in 3\angs\ spectral regions. At the bottom is the Sun: two scans of the 
center of the solar disk are superimposed. At the very top, the strongest lines 
in the solar spectrum calculation are identified as described in the text. 
The same mid-UV normalization constants as in Fig.\ 1 are adopted for the top 
nine stars.  $a$) 2627.5 -- 2630.5\angs. $b$) 2642.5 -- 2645.5\angs. $c$) 2900 -
- 2903\angs. $d$) 3082.5 -- 3085.5\angs. $e$) 3100 -- 3103\angs.

\figcaption[fig4a-f.eps]{\label{hd9.wav}} Comparisons as in Fig.\ 1 are shown 
for 
optical spectral regions containing magnesium or Balmer lines in eight of the 
stars plus the Sun. 
Each observed spectrum (heavy line) has been normalized to unity as 
described in the text. Line identifications as described for Fig.\ 3 are shown 
for the 
strongest lines in the solar calculation. $a$) \mgj. $b$) Low-excitation \mgi. 
$c$), $d$) The \mgb\ lines. $e$) High-excitation \mgi. $f$) \Halpha. 
In this panel only, the third plot from the top represents a non-optimal choice 
of model parameters. It illustrates the degradation of the fit to \Halpha\ when 
a \teff\ is chosen 100\kel\ cooler than that of the plot immediately below.
\clearpage

\begin{deluxetable}{rcccccccccccccccccccc}
\tablenum{1}
\tablecolumns{20}
\tabletypesize{\scriptsize}
\tablecaption{\bf Stellar Observations, Model Parameters, and Effective 
Temperatures}
\tablewidth{0pt}
\rotate
\tablehead{
\multicolumn{21}{r}{$\Leftarrow$~~~~~~~~~~~~~~~~Observed
~~~~~~~~~~~~~~~~~$\Rightarrow$~~Mid-UV~~~$\Leftarrow$~~~~~~~~Observed $-$ 
Model~~~~~~$\Rightarrow$~~~~~$\Leftarrow$~~~~~~~~~~Model~~~~~~~~~~~~$\Rightarrow
$~~~~~K93~~~~C+94~~~A+96~~~G+96~~~BL98~~~F00}\\
\colhead{HD} & 
\colhead{$V$} & 
\colhead{\bmv} & 
\colhead{\umb} & 
\colhead{\bmy} & 
\colhead{$\pi$} & 
\colhead{Ratio} & 
\colhead{$V$} & 
\colhead{\bmv} & 
\colhead{\umb} & 
\colhead{\bmy} &
\colhead{\teff} & 
\colhead{\logg} & 
\colhead{\feh} & 
\colhead{\vt} & 
\colhead{\teff} & 
\colhead{\teff} & 
\colhead{\teff} & 
\colhead{\teff} & 
\colhead{\teff} & 
\colhead{\teff}

}
\thispagestyle{empty}
\startdata
 19445 & +8.06 & 0.45 & $-0.25$ &  0.349 & 25.85& 1.37 & $-0.04$ & $-0.01$ & 
$-0.01$ & $-0.005$ & 6050 & 4.5 & $-2.00$ & 1.0 & 6007   & 5842   & 6050   & 6066   
& \ldots & 5825\\
 84937 & +8.33 & 0.39 & $-0.22$ &  0.303 & 12.44 & 1.21 &  +0.09  & $-0.01$ & 
$-0.01$ & $-0.011$ & 6300 & 4.0 & $-2.20$ & 1.5 & 6314   & 6220   & \ldots & 6350   
& 6202   & 6375\\
 94028 & +8.22 & 0.47 & $-0.18$ &  0.343 & 19.23 & 1.07 &  +0.25  &  +0.00  &  
+0.00  & $-0.005$ & 6050 & 4.2 & $-1.40$ & 1.3 & 6047   & 5898   & \ldots & 6060   
& 5998   & 5900\\
106516 & +6.11 & 0.45 & $-0.14$ &  0.317 & 44.34 & 0.99 & $-0.09$ & $-0.01$ & 
$-0.03$ & $-0.004$ & 6250 & 4.3 & $-0.65$ & 1.0 & 6221   & 6118   & 6208   & 6267   
& 6233   & 6200\\
114762 & +7.31 & 0.53 & $-0.06$ &  0.365 & 24.65 & 0.96 &  +0.30  & $-0.02$ &  
+0.01  & $-0.008$ & 5850 & 4.0 & $-0.90$ & 1.0 & \ldots & 5790   & \ldots & 5941   
& 5904   & 5800\\
128167 & +4.46 & 0.36 & $-0.08$ &  0.254 & 64.66 & 0.92 & $-0.05$ &  +0.01  & 
$-0.01$ & $+0.008$ & 6850 & 4.3 & $-0.35$ & 2.0 & \ldots & \ldots & 6707   & 6734   
& 6737   & \ldots\\
128620 & +0.00 & 0.67 &  +0.23  & 0.438 & 742.2 & 0.46 &  $-0.06$ &  +0.00  &  
+0.01 & $+0.037$ & 5800 & 4.3 &  +0.15  & 1.0 & \ldots & \ldots & \ldots & 
\ldots & \ldots & \ldots\\
184499 & +6.62 & 0.58 &  +0.00  &  0.390 & 31.29 & 1.11 &  +0.02  & $-0.02$ & 
$-0.02$ & $+0.001$ & 5750 & 4.0 & $-0.60$ & 1.2 & 5734   & 5726   & 5750   & 5750   
& 5773   & 5700\\
201891 & +7.37 & 0.51 & $-0.17$ &  0.358 & 28.26 & 0.92 &  +0.40  & $-0.02$ & 
$-0.07$ & $-0.008$ & 5900 & 4.1 & $-1.00$ & 1.0 & 6014   & 5817   & \ldots & 5974   & 5918   & 5825\\
Sun   & $-26.7$ & 0.65~ & +0.13 & \ldots & \ldots & 0.91 & \ldots  & $-0.01$ & $-0.04$ & \ldots  
& 5775 & 4.44 & +0.00 & 1.0 & \ldots & \ldots & \ldots & \ldots & \ldots & \ldots\\

\enddata

\tablecomments{Units: $\pi$\ in $10^{-3}$\arcsec; \teff\ in~\kel; \vt\ in \kms. Sources: 
For all stars but \hd 128620 = \acena, $V$, \bmv, and \umb\ from \citet{car94}; \bmy\ 
from \citet{kin93}, except \bmy\ for \hd 114762, \hd 128167, and \hd 201891 from SIMBAD; $\pi$\ from SIMBAD.
\hd 128620 = \acena: $V$, \umb, \bmy, and $\pi$\ from SIMBAD; \bmv\ from \citet{fur90}. 
All solar observations from \citet{all73}. Other \teff\ sources: K93, \citet{kin93}; C+94, 
\citet{car94}; A+96, \citet{alo96}; G+96, \citet{gra96}; BL98, \citet{bla98}; F00, \citet{ful00}.
}

\end{deluxetable}

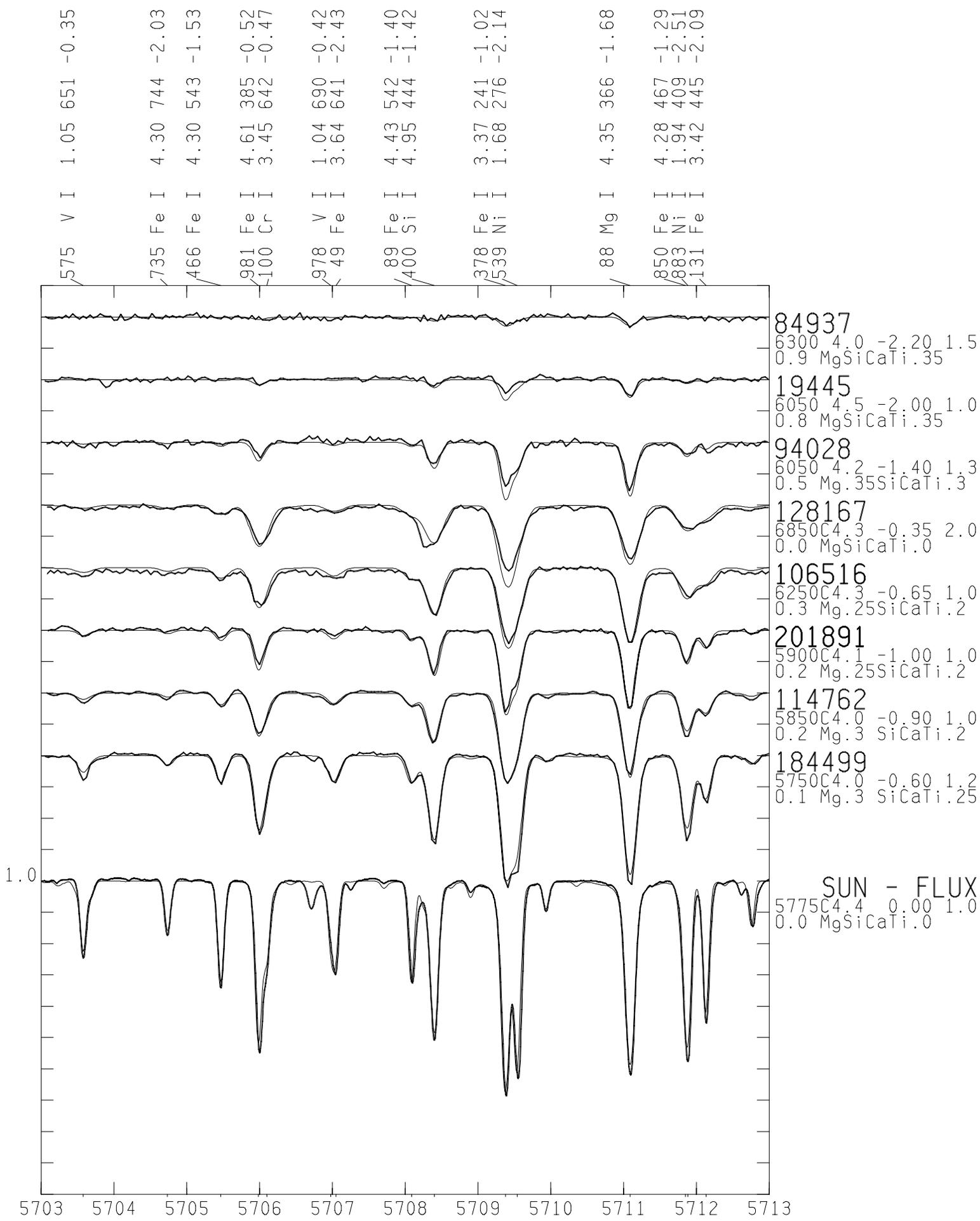
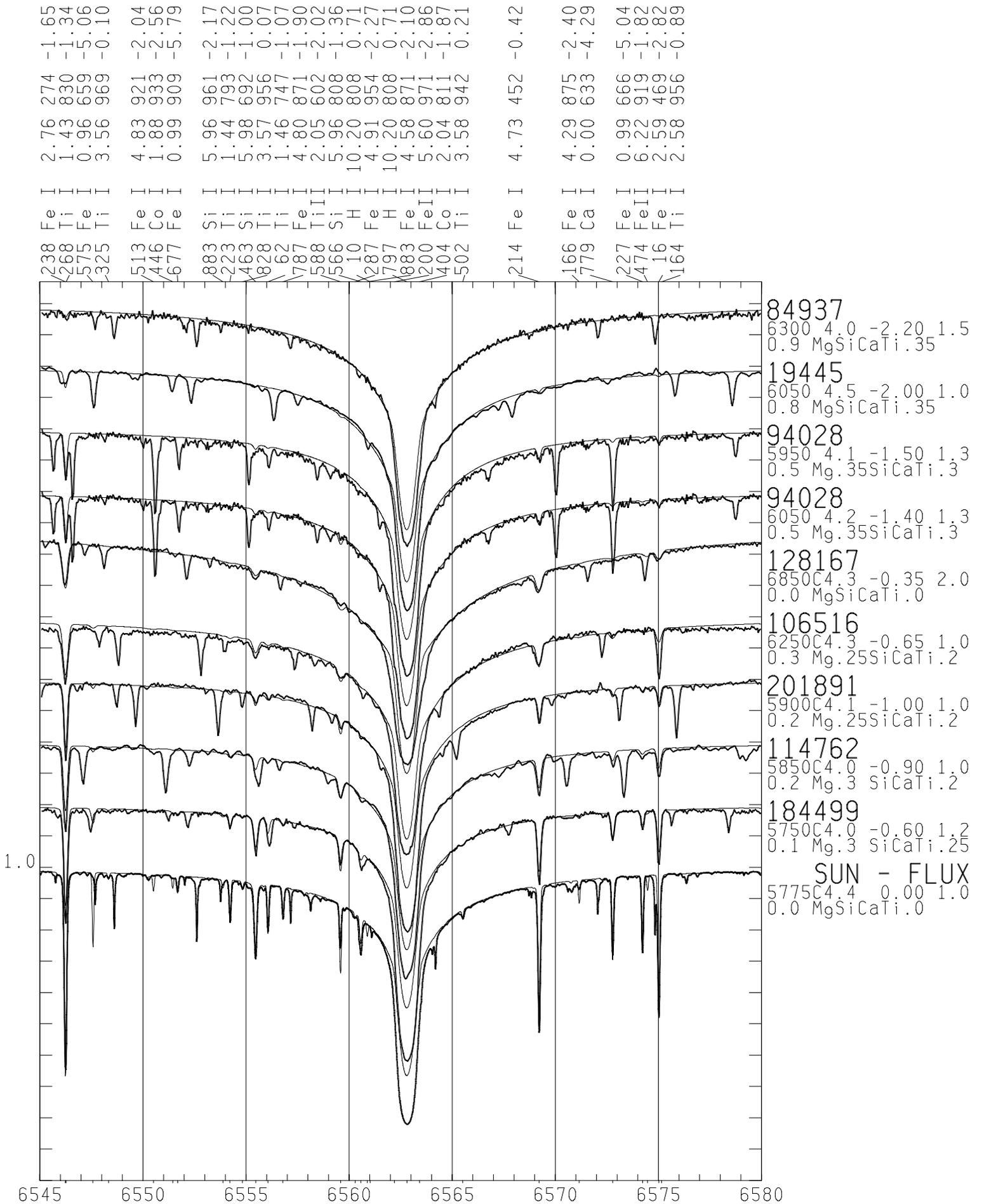
\end{document}